\documentclass[aps,superscriptaddress,secnumarabic]{revtex4}
\usepackage{epsfig,graphics,amssymb,amsmath}

\begin{document}
\title[Command Option Summary]{Adhesion Transition of Flexible Sheets}
\author{Arthur A. Evans}
\email{aevans@physics.ucsd.edu}
\affiliation{Department of Physics,}
\author{Eric Lauga}
\email{elauga@ucsd.edu}
\affiliation{Department of Mechanical and Aerospace Engineering,\\ University of California San Diego, 9500 Gilman Drive, La Jolla CA 92093.}
\date{\today}
\begin{abstract}
Intermolecular forces are known to  precipitate adhesion events between solid bodies.  Inspired by a macro-scale experiment showing the hysteretic adhesion of a piece of flexible tape over a plastic substrate,  we develop here a model of far-field dry adhesion between two flexible sheets interacting via a power-law potential.  We show that phase transitions from unadhered to adhered states occur as dictated by a dimensionless bending parameter representing the ratio of interaction strength to bending stiffness.  The order of the adhesion transitions, as well as their hysteretic nature,  is shown to depend on the form of the interaction potential between the flexible sheets.  When three or more sheets interact, additional geometrical considerations determine the hierarchical  or  sequential nature of the adhesion transitions.
\end{abstract}
\maketitle

\section{Introduction}

As fabrication technology and nanoscale engineering increase in complexity, it becomes vital to understand small-scale interactions between material components.  Surface-tension mediated forces play a large role in self-assembly, not only at the macro-scale \cite{elastocap}, but also for micro-electromechanical and nano-electromechanical structures (MEMS and NEMS), and as such, a large amount of work has been done in studying the adhesive forces involved \cite{NEMS,hierarch,caprise}.  Carbon nanotubes (CNTs) have attracted significant attention since they were discovered to exhibit novel electrical and mechanical properties, and it has been found that CNTs can adhere to each other under the influence of capillary forces \cite{nanoforest,nanotrib,CNTtension}.  At these scales, fluid-regulated forces are not the only factors that must be examined. Dispersion (or van der Waals) forces may become more important than at larger scales, and the microscopic intermolecular forces of extended media start to have a macroscopic effect on structural stability \cite{langbein,intermolec}.

In addition to progress in nanotechnology, many biological systems also display adhesion phenomena whose origins can be traced to intermolecular forces.  Geckos are known to adhere to smooth surfaces, without any liquid interface. The microscopic arrays of hairs, or setae, on the base of the gecko foot are therefore believed to be the source of such effective dry adhesion \cite{gecko,CNTgecko,gecko_hierarch,spatula,VDWsetae}.  In cellular biology, cytoskeletal morphogenesis is regulated by complex biopolymer networks: Series of long, thin, elastic filaments that form a scaffolding for eukaryotic cells.    Mechanical properties of macromolecules such as actin filaments or DNA can be measured by force or deflection analysis at small scales, and polymers adsorbed onto a surface or  ``zipped'' to another molecule can be peeled apart by applying optical tweezers or other external pulling forces \cite{DNAadhesion,DNAhysteresis,polymerunzip,bottom_up_cell}.

Most of the research into interactions between materials at these scales involve close-range, contact, and sometimes capillary forces, and this is the limit considered by many models and experiments to date \cite{DNAadhesion,DNAhysteresis,polymerunzip}.   However, long-range forces due to fixed charge distributions,  polar, or even non-polar interactions  can lead to adhesion events if the right conditions are satisfied.  
Previous work has characterized the van der Waals attraction between thin flexible objects, both theoretically \cite{oyhar,tweezers} and experimentally  \cite{peeling}.  
In this paper, we aim to develop an understanding of the physical mechanisms by which  long-range interaction forces compete with elasticity in the adhesion of thin, flexible structures.  We first introduce and motivate the prototypical system of interest using a macro-scale  experiment showing the hysteretic adhesion of a piece of flexible tape over a plastic substrate. We then   develop a model of far-field dry adhesion between two elastic, slender sheets interacting via a power-law potential, and study numerically their  relative adhesion.  We uncover that phase transitions from unadhered to adhered states occur as dictated by a dimensionless bending parameter representing the ratio of interaction strength to bending stiffness, as well as the form of the interaction potential between the flexible sheets. We then generalize our model in order to study the interactions between several sheets, and show that additional geometrical considerations determine the hierarchical  or  sequential nature of the adhesion transitions in that case.

\section{Macro-scale experiment}
\label{macroexp}

\begin{figure}[t]
\includegraphics[width=.45\textwidth]{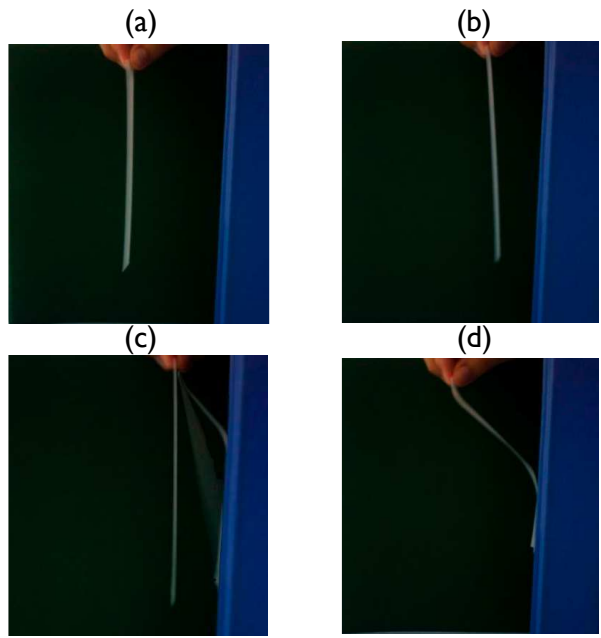}
\caption{(Color online) \label{pix} An example of an adhesion transition between flexible sheets. A piece of adhesive tape is charged electrostatically and then moved slowly towards an uncharged surface, with the adhesive side turned away from the surface.  (a): The charged tape is held far from from the surface and no noticeable bending occurs; 
(b): Weak bending is exhibited just before the critical transition point; 
(c): At a critical distance the tape moves rapidly towards the surface; 
(d): Moving the tape away from the wall back to its original position shows hysteretic behavior in the shape of the tape.}
\end{figure}

An example of adhesion transition between between elastic bodies due to long-range  interactions  may be demonstrated using everyday materials, namely a piece of adhesive tape and a plastic substrate.  As shown in Fig. \ref{pix}, this tape can be shown to exhibit complex adhesion properties. The tape is first given a static charge distribution by applying it to a piece of plastic and then removing it swiftly.  The tape is initially suspended at the distance shown in Fig.~\ref{pix}a, sticky side away from an uncharged substrate.  As the suspension distance is slightly decreased (Fig.~\ref{pix}b), the tape becomes weakly attracted to the surface.  At a critical distance (Fig.~\ref{pix}c), the attraction suddenly pulls the tape completely to the surface, where it lays flat along the majority of the substrate.  As the tape is pulled away from the surface, the shape exhibits hysteresis (Fig.~\ref{pix}d).  As the top of the tape is returned to its initial position, the shape remains stuck to the surface, even past the distance where it first adhered.   

This simple macro-scale experiment allows us to introduce some qualitative features  of the adhesion transition, namely a competition between bending and long-range interaction, a sharp transition in shape, and hysteresis.  We present below a numerical approach to quantify the behavior of similar, but more general, systems.  Note that there are other characteristics of the macroscopic experiment that we will not attempt to model in this work, namely the presence of dynamic effects and force due to gravity.

\section{Theory}

\subsection{Setup}

\begin{figure}[t]
\includegraphics[width=.4\textwidth]{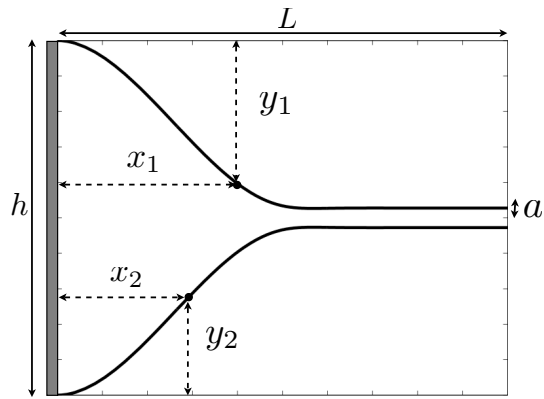}
\caption{\label{shape} A schematic representation of a two-dimensional cross-section for a  system of two flexible sheets (see text for notation).  Our analysis will be limited to the regime where $\epsilon=h/L \ll 1$.}
\end{figure}

The system that we study is displayed schematically in Fig.~\ref{shape}.  Two sheets of length $L$, thickness $a\ll L$, and width $d$ (not shown) are clamped at their left-most edges, separated by a distance $h$, and free to interact along their lengths.   We assume the deformations to be two dimensional, and describe each sheet by the vertical deformation of its centerline, denoted $y_i$, with $i=1,2$. While DNA and other semiflexible polymers can become kinked, looped, and otherwise knotted, this study will be limited to the case where the length ratio $\epsilon=h/L \ll 1$, i.e. the long-wavelength limit.

Under these assumptions the total energy of the system is given by
\begin{eqnarray}
\label{energy}
E=\frac{1}{2}B_1\int_0^L{y_1^{\prime\prime}(x_1)^2dx_1}+\frac{1}{2}B_2\int_0^L{y_2''(x_2)^2dx_2}+ \nonumber \\\int_0^L{\int_0^L{V[x_1,x_2,y_1(x_1),y_2(x_2)]dx_1}dx_2},
\end{eqnarray}
where $B_i$ is the bending modulus of the $i^{th}$ sheet, and $x_i$  is the horizontal distance. 
The function V describes the interaction potential energy density between the two sheets, as yet unspecified (note that the integration along the widths of the sheets has already been performed formally in $V$).  Extremizing this functional yields mechanical equilibrium, as shown by the following system of coupled integro-differential equations with boundary terms
\begin{eqnarray}
\label{nonlocal}
& \displaystyle B_1y_1^{\prime\prime\prime\prime}+\int_0^L{dx_2\frac{\partial V}{\partial y_1}}=0, \label{systA} \\
 & \displaystyle B_2y_2^{\prime\prime\prime\prime}+\int_0^L{dx_1\frac{\partial V}{\partial y_2}}=0,\\
\displaystyle & \displaystyle y_1^{\prime\prime\prime}\delta y_1\Big{|}_0^L=0\,\,\,\,\,\,\,\, y_1^{\prime\prime}\delta y_1^{\prime}\Big{|}_0^L=0,\\
\displaystyle &y_2^{\prime\prime\prime}\delta y_2\Big{|}_0^L=0\,\,\,\,\,\,\,\,y_2^{\prime\prime}\delta y_2^{\prime}\Big{|}_0^L=0. \label{systD}
\end{eqnarray}
The boundary conditions are set by the physical conditions of the sheets.  While there are many possible cases that could be examined, we will consider the common physical scenario in which the sheets are fixed and clamped on the left ($y_1(0)=h,\, y_2(0)=0,\,y_i^{\prime}(0)=0$, $i=1,2$) and the right edge of the sheets are force- and moment-free ($y_i^{\prime\prime}(L)=y_i^{\prime\prime\prime}(L)$=0, $i=1,2$).

The potential $V$ can be chosen to describe the physical mechanism responsible for the adhesion between the sheets, \cite{hierarch,polymerunzip,gecko}. In this paper, we are considering a general long-range potential of the form  V $\sim 1/r^n$, where $r=\sqrt{(x_1-x_2)^2+(y_1(x_1)-y_2(x_2))^2}$, and $n$ is a positive integer.  More specifically,  we set
\begin{equation}
\label{potential}
V=\sum_{n=1}^{N}\frac{A_nW\sigma^{n}}{[(x_1-x_2)^2+(y_1(x_1)-y_2(x_2))^2]^{n/2}},
\end{equation}
where $\sigma$ is the van der Waals-like radius, $W$ is the strength of the interaction, and $N$ is the number of singular modes.  The sign of $A_n$ determines whether the interaction is attractive or repulsive.  We will examine the more specific form of this general potential where only two terms remain, an attractive term $n=p$ with $A_p=-1$ and a repulsive term $n=q$ with $A_q=+1$.  This is the familiar Lennard-Jones-like potential that is used to model intermolecular interactions \cite{intermolec}.  We will also work with the case that $\sigma>a$, so no ``true" contact between the sheets will occur.  In a related study, Oyharcabal and Frisch \cite{oyhar} use a van der Waals-like medium range potential with values of $p=3$ and $q=9$ to model the attraction between a thin filament and a nonpolar substrate.  Other examples include the van der Waals interaction between two filaments ($p=6$, $q=12$), polarized attraction between two sheets ($p=2$, $q>p$), Coulombic attraction ($p=1$, $q>p$), and many others (see Ref.~\cite{intermolec} for a review).  In fact, a surface with an arbitrary charge distribution can be represented by a standard multipole expansion, and in a suitable far-field regime a charged polymer or conducting elastic sheet can be modeled by this potential as well.  Very generally, by specifying the values of $p$ and $q$, any number of potential interactions can be represented, except in the rather exceptional cases in which a power-law potential model is insufficient.

\subsection{Dimensionless Equations}
 The system described by Eqs.~\eqref{systA}-\eqref{systD} is non-dimensionalized by scaling the vertical displacements by $h$, and horizontal distances by $L$.  In what follows variables are understood to be dimensionless.   In that case, Eqs.~\eqref{systA}-\eqref{systD} become
\begin{eqnarray}
\label{bleh}
\label{7}&y_1^{\prime\prime\prime\prime}(x_1)+\Pi_{p,1}I(x_1,y;p,2)-\Pi_{q,1}I(x_1,y;q,2)=0,\\
\label{8}&y_2^{\prime\prime\prime\prime}(x_2)-\Pi_{p,2}I(x_2,y;p,1)+\Pi_{q,2 }I(x_2,y;q,1)=0,\\
&y_1(0)=1, \,\,\,\,\,\,y_1'(0)=y_1''(1)=y_1'''(1)=0,\\
\label{10}&y_2(0)=0, \,\,\,\,\,\,y_2'(0)=y_2''(1)=y_2'''(1)=0,
\end{eqnarray}
where $\Pi_{p,i}=p\sigma^{p}L^{3-p}W/B_i$ is a dimensionless quantity, and where  we have defined the integral $I(x_i,y;\alpha,k)$ as
\begin{equation}
\label{integral}
I(x_i,y;\alpha,k)=\\ \int^1_0{\frac{y\,dx_k}{\left[(x_i-x_k)^2+\epsilon^2y^2 \right]^{\frac{\alpha}{2}+1}}},
\end{equation}
with $y=y_1(x_1)-y_2(x_2)$.

\subsection{Asymptotics}
We now take advantage of the long wavelength approximation ($\epsilon \ll 1$) to simplify the integrals of the form 
$I(x_i,y;\alpha,k)$.  Introducing the substitution $x_1-x_2=\epsilon\xi$  we obtain
\begin{equation}
\label{integrals}
I(x_1,y;\alpha,2)=\frac{1}{\epsilon^{\alpha+1}}\int^{\frac{1-x_1}{\epsilon}}_{\frac{-x_1}{\epsilon}}{\frac{y_1(x_1)-y_2(x_1-\epsilon\xi)}{(\xi^2+[y_1(x_1)-y_2(x_1-\epsilon\xi)]^2)^{\frac{\alpha}{2}+1}}}d\xi
.\end{equation}

Expanding to leading order in $\epsilon$,
\begin{gather}
I(x_1,y;\alpha,2)=\frac{1}{\epsilon^{\alpha+1}}\int^{\infty}_{-\infty}\frac{{y_1(x_1)-y_2(x_1)d\xi}}{(\xi^2+[y_1(x_1)-y_2(x_1)]^2)^{\frac{\alpha}{2}+1}}  + o\left(\frac{1}{\epsilon^{\alpha+1}}\right)\\
=\frac{1}{\epsilon^{\alpha+1}[y_1(x_1)-y_2(x_1)]^{\alpha}}\int^{\infty}_{-\infty}\frac{du}{(1+u^2)^{\frac{\alpha}{2}+1}} + o\left(\frac{1}{\epsilon^{\alpha+1}}\right) \\
=\frac{\sqrt{\pi}}{\epsilon^{\alpha+1}[y_1(x_1)-y_2(x_1)]^{\alpha}}\frac{\Gamma(\frac{1+\alpha}{2})}{\Gamma(1+\frac{\alpha}{2})} + o\left(\frac{1}{\epsilon^{\alpha+1}}\right),\label{end}
\end{gather}
where we have used $u=\xi/[y_1(x_1)-y_2(x_1)]$.  The other integrals in Eqs.~\eqref{7}-\eqref{8} are evaluated similarly.  Physically, Eq.~\eqref{end} expresses the fact that, in the long wavelength limit $L\gg h$, each sheet see the other one as being locally flat, and therefore at leading order the integration along the horizontal direction can be performed first.

\subsection{Identical sheets}
Having derived above the general system of equations for two interacting sheets, we now consider the simplified case where the sheets are identical.  Defining  $z(x)$ as the distance between the sheets, $z(x)=y_1(x)-y_2(x)$, Eqs.~\eqref{7}-\eqref{10}  become
\begin{eqnarray}
\label{asympeom}
\label{z1}&\displaystyle z^{\prime\prime\prime\prime}+\Omega\left(\frac{1}{z^p} -\frac{\beta}{z^{q}}\right)=0,\\
&z(0)=1, \,\,\,\,\,\,z'(0)=0,\\
\label{z3}&z''(1)=0,\,\,\,\,\, z'''(1)=0,
\end{eqnarray} 
where $\Omega=(2WJ_pL^3\sigma^p)/(Bh^p\epsilon)$, $\beta=(\sigma/h)^{q-p}J_q/J_p$ and $J_{\alpha}=\Gamma((1+\alpha)/2)/\Gamma(1+\alpha/2)$.
Note that the divergent behavior $z=0$ is prohibited thanks to the repulsive part of the potential in Eq.~\eqref{asympeom}.
The dimensionless quantity $\Omega$, which we refer to as the bending parameter, is a measure of the relative importance of the interaction forces to the elastic forces, while $\beta$ is a dimensionless van der Waals-like radius with a numerical prefactor. Hence the symmetric system is completely described by the four parameters $\{\Omega, \beta, p,  q\}$.

\subsection{Numerics}

The symmetric nonlinear system described by Eqs.~\eqref{z1}-\eqref{z3} is solved numerically on an adaptive grid to an absolute error tolerance $10^{-12}$, along with a standard Newton-Raphson shooting method using MATLAB. A continuation scheme in $\Omega$ and $\beta$ allows efficient computation of nearby systems.  This treatment is similar to that followed in Ref.~\cite{oyhar}.

\section{Adhesion transition}

\subsection{Main result}

\begin{figure*}[t]
\includegraphics[width=.8\textwidth]{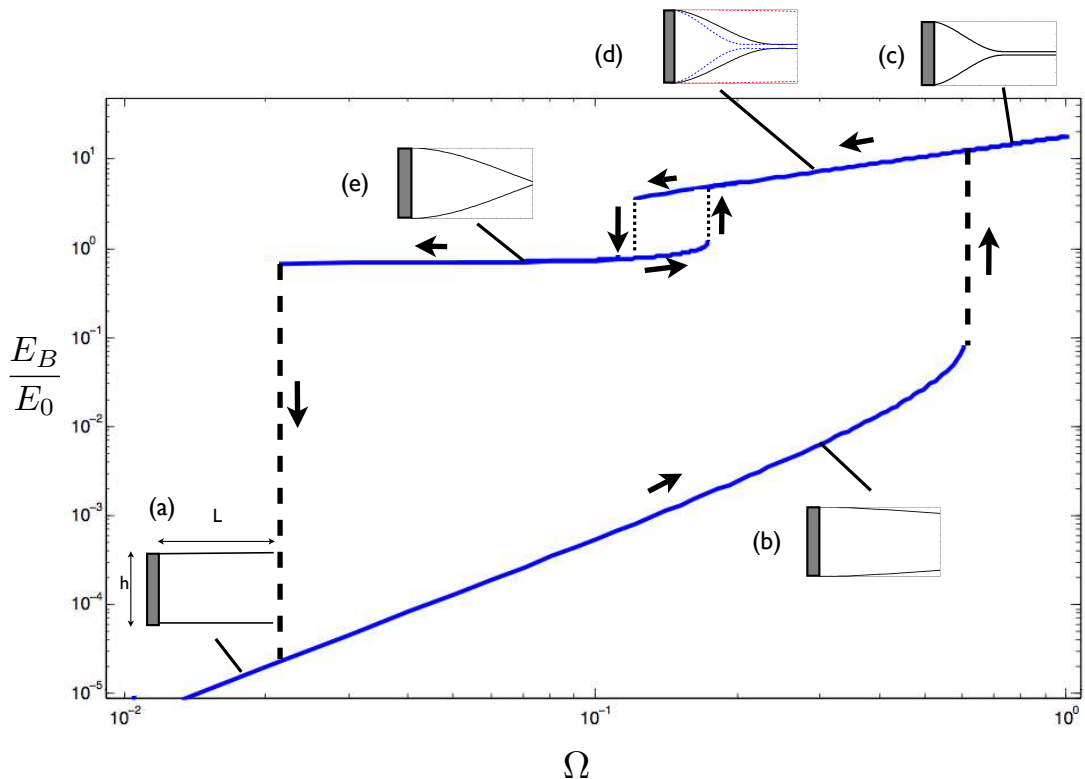}
\caption{(Color online) \label{hysteresis} Dimensionless bending energy $E_B/E_0$ as a function of the bending parameter, $\Omega$, with $p=3$, $q=9$ and $\beta=0.05J_9/J_3$.  Here $E_0$ is a typical bending energy,  $E_0=Bh^2/L^3$.  Representative shapes of the two interacting sheets are shown in the different regions. In Fig.~\ref{hysteresis}a, the sheets are essentially straight.  Fig.~\ref{hysteresis}b shows a slightly bent state due to weak attraction between the sheets.  Fig.~\ref{hysteresis}c: Past a critical value of $\Omega$, the sheets abruptly adhere to one another.  As $\Omega$ is decreased, the sheets retain their adhered character, although the shapes change, as seen in Fig.~\ref{hysteresis}d (dashed lines indicate shape from Fig.~\ref{hysteresis}c).  There is also another sharp transition as $\Omega$ is decreased even more, and the sheets detach into a bent arc-like shape (Fig.~\ref{hysteresis}e).  As $\Omega$ is decreased further still there is a final sharp transition back to the original weakly attracted shapes shown in Fig.~\ref{hysteresis}a.}
\end{figure*}

The main result of this paper is illustrated in Fig.~\ref{hysteresis}. For the particular values  $p=3$, $q=9$ and $\beta=0.05J_9/J_3$, we display the nondimensional bending energy of the sheets $E_B$ --- {\it i.e.} the sum of the first two terms in Eq. (\ref{energy}) --- as a function of the bending parameter $\Omega$. Although here we have chosen values for $\{p,q,\beta\}$, the results are similar for other values, with some possible  qualitative differences  highlighted in the sections below.

The sudden shape changes, quantified by the bending energy,   are reminiscent of the behavior observed experimentally in  \S\ref{macroexp}. For $\Omega \ll 10^{-2}$, the two sheets are essentially free-standing, as indicated in Fig.~\ref{hysteresis}a.  As $\Omega$ is increased the sheets are attracted weakly to one another, resulting in a small shape change (Fig.~\ref{hysteresis}b).  At a critical value of $\Omega$, the bending energy jumps discontinuously and the sheets abruptly snap together (Fig.~\ref{hysteresis}c).  As  $\Omega$ increases further, the sheets become more tightly bound, with the unclamped portion becoming smaller.  If $\Omega$ is then decreased, the system exhibits hysteretic behavior, with the sheets remaining adhered as shown in Fig.~\ref{hysteresis}d (dashed lines indicate the previous shape from Fig.~\ref{hysteresis}c).  As $\Omega$ is decreased further still there is another discontinuity in the energy, and the sheets once again take on a qualitatively different shape, arc-like, as displayed in Fig.~\ref{hysteresis}e.  Remarkably, there exists a second (smaller) hysteresis loop on this branch of the energy profile.  The jump in the energy at this second hysteresis corresponds to a large change in the contact between the sheet end points and the slope of the sheets.  Finally, for decreasing $\Omega$, the sheets return to the positions shown by Fig.~\ref{hysteresis}a via another sharp transition.

\begin{figure}[t]
\includegraphics[width=.4\textwidth]{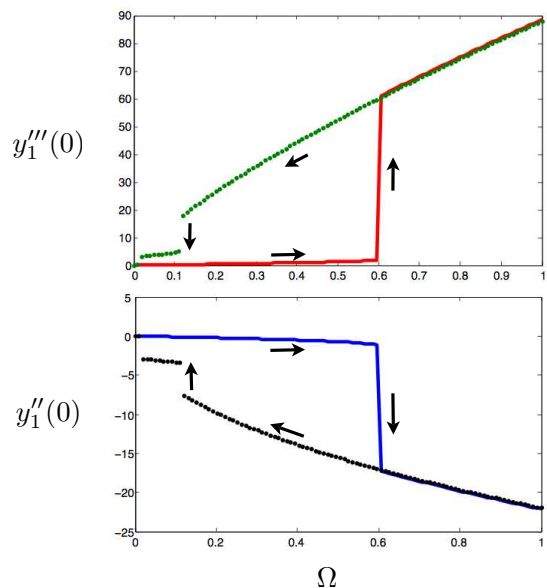}
\caption{(Color online) \label{ics} Dimensionless force, $y_1'''(0)$, and moment, $y_1''(0)$, on the left side of the first sheet as a function of the bending parameter, $\Omega$, for values of $p=6$, $q=12$, and $\beta=0.15J_{12}/J_6$.  Note the lack of a second hysteresis loop for these values of the parameters.}
\end{figure}

We also plot in Fig.~\ref{ics} the dimensionless force, $y_1'''(0)$, and moment, $y_1''(0)$, necessary to apply  to the left edge of the first sheet to maintain it clamped.  For the values of $\{p,q,\beta\}$ considered (6, 12, and $0.15J_{12}/J_6$, respectively), the system models two filaments interacting via van der Waals forces.  Much like the shapes themselves, the forces  and moments undergo  sharp transitions and exhibits hysteresis.
Note that if the sheets were free to interact they would adhere along their entire length, and a force would need to be applied to one end in order to peel them apart. In essence the same effect is seen in our system.  Hysteresis is known to occur in the strong loading of cantilevers \cite{bifurcate}, and recent experimental investigations into the peeling of CNTs from a substrate have reported results  qualitatively similar  to ours \cite{peeling}.

\subsection{Phase transition and physical quantities}

\begin{figure}[t]
\includegraphics[width=.8\textwidth]{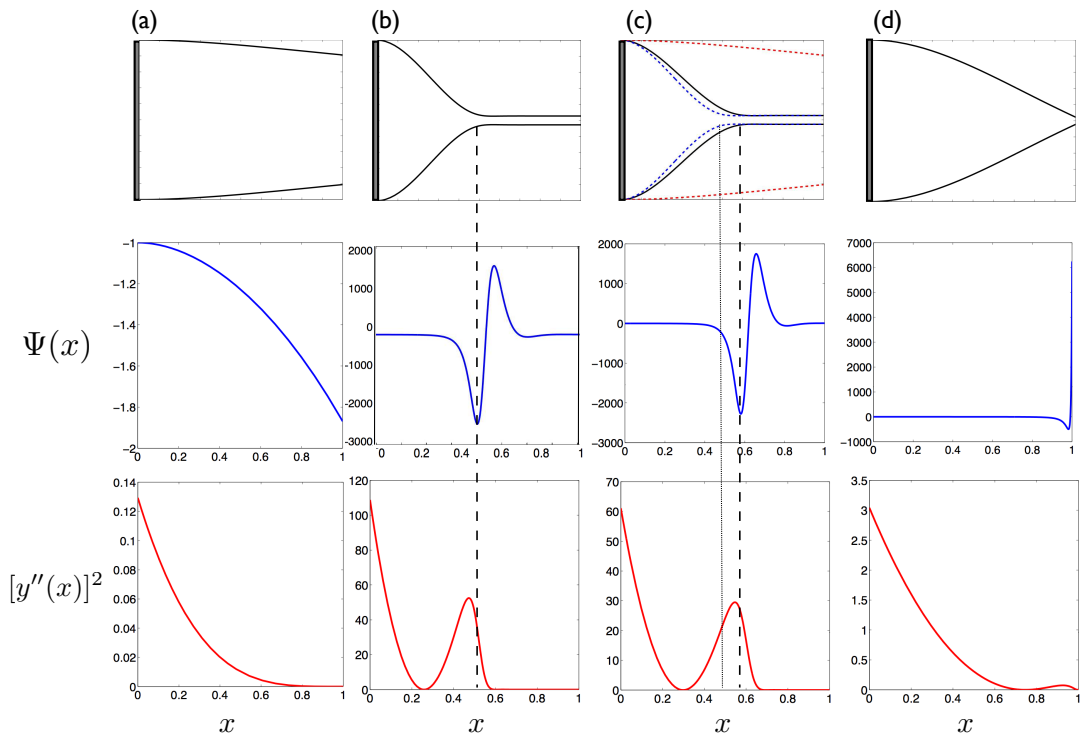}
\caption{(Color online) \label{density} Bending energy density, $[y''(x)]^2$, and interaction energy density, $\Psi(x)$, for various sheet shapes (top) along the hysteresis loop for $p=3$, $q=9$, and $\beta=0.05J_9/J_3$: (a) Weak bending ($\Omega=0.5$), with ends of the sheets approximately straight (zero bending energy); (b):  Tightly clamped configuration ($\Omega=0.89$) with local minimum in the interaction energy density denoting the end of the unclamped region.  The local minimum in the interaction energy is in the same vicinity, but not at the same point; (c): Hysteretic clamped shape ($\Omega=0.5$), with vertical dashed line indicating  the position of the local maximum of bending energy density, and the vertical dotted line indicates where this local maximum was in Fig.~\ref{density}b. The red (outer horizontal dashed) line in the upper inset denotes the weakly bent shape from (a), while the blue (inner horizontal dashed) line denotes the shape from (b). (d): Arc-shape $(\Omega=0.11$), with a notable small local maximum in $\Psi(x)$ near the end point indicating localized adhesion. }
\end{figure}

By viewing the shape-change as a phase transition, where the control parameter  is  $\Omega$ instead of temperature, we can borrow several concepts from statistical physics in order to further characterize  our model system.  
The natural order parameter to assign is the distance between the sheets, $z(x)$, as $z(x)=1$ denotes totally unadhered sheets and $z(x)=\beta^{1/(q-p)}$ corresponds to complete adhesion (see Eq.~\ref{asympeom}). 
A natural analogy exists between the energy functional given by Eq.~\eqref{energy} and a one-dimensional magnetic system with two-component spin subject to an external field \cite{stretch_fil}.  In our system, there is an energy penalty associated with deforming the sheets (analogously, misaligning spins), and there is an interacting field that acts to order the system (analogously, the external magnetic field).  It is  known that even at zero temperature, the magnetic system displays a phase transition at a critical value of the ordering field (except in the thermodynamic limit of the sheet length $L\rightarrow\infty$), and as such we could expect such behavior from our system as the relative field strength (i.e. $\Omega$) is increased.

In analogy to the external field of the magnetic system, we define an interaction energy density given by
$\Psi(x)=\Omega\left(-{1}/{z^p}+{\beta}/{z^q}\right)$, and discuss the qualitative changes that govern the phase behavior of the system by studying the minima in the free energy  (see Ref.~\cite{kardar} for a textbook treatment).  When a local minimum appears or disappears along the length of the sheets we can expect a change in shape, and whether this change is dramatic or smooth will correspond to a first- or second-order phase transition (first-order when $\partial E/\partial\Omega$ is discontinuous, second-order when $\partial^2E/\partial\Omega^2$ is discontinuous).

In Fig.~\ref{density}, we display a representative sampling of the energy densities (bending and interaction energies) with their associated shapes, for $p=3$, $q=9$, and $\beta=0.05J_9/J_3$.  We see  qualitatively different energy densities, confirming the transitions between three different phases. In Fig.~\ref{density}a, the sheets store little elastic energy and are only weakly attracted.  Past the critical adhesion point, the shape as displayed in Fig.~\ref{density}b now shows a large energetic favorability from the interaction force, with large deformation energy penalties on the left edges of the sheets and at the end of the un-adhered length.   As the bending parameter is increased further,  the spatial location of the energy minima shifts, as displayed by Fig.~\ref{density}c.  The length of the adhered region (or domain wall), $\delta$, increases with $\Omega$  as $\delta \sim \Omega^{-1/4}$, as expected from the boundary layer scaling arising from  Eq.~\eqref{asympeom}. This scaling is confirmed by our numerical simulations (not reproduced here). Decreasing $\Omega$, and coming down on the hysteresis loop, at a lower  critical value of $\Omega$ the adhered sheets  become arc-like, and the second local minimum in $\Psi(x)$ disappears (see Fig.~\ref{density}d).

\subsection{General Behavior} 

\begin{figure*}[t]
\includegraphics[width=.8\textwidth]{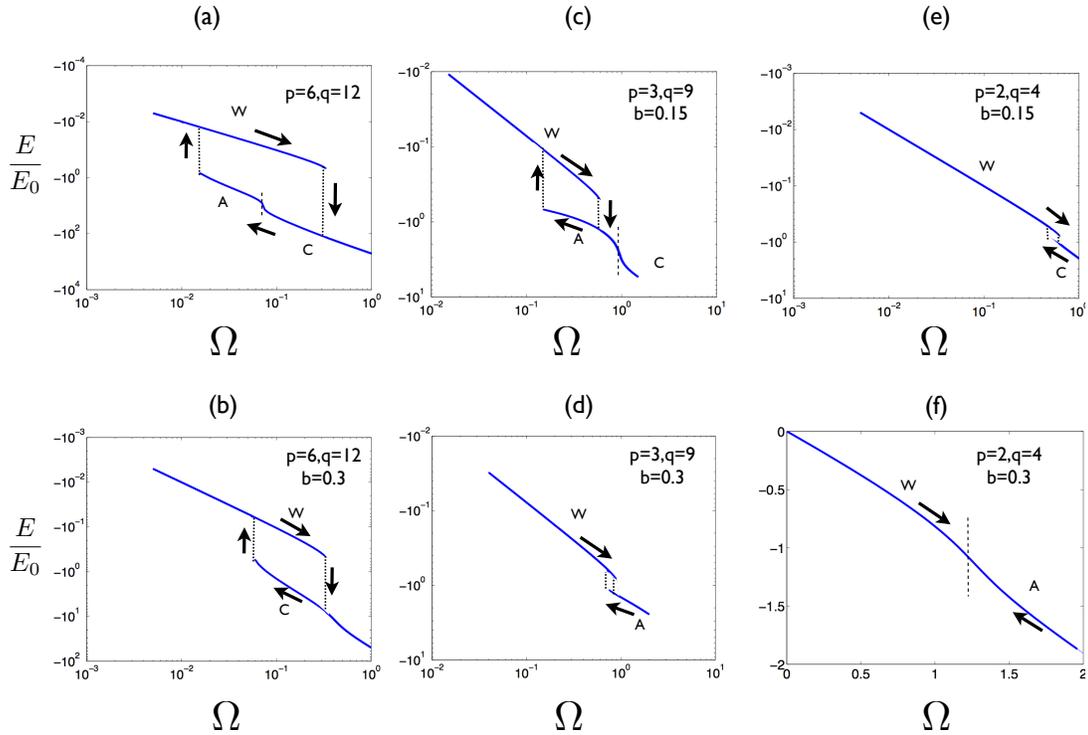}
\caption{(Color online) \label{energies} A comparison of total energies for the system (bending + attraction energy) for different values of $p$, $q$, and $\beta$.  For convenience we have introduced the parameter $b=\beta J_p/J_q$.   (a): For $p=6$ and $q=12$, these values model the attraction between two thin non-polar filaments.  Dotted lines denote sharp phase transitions, while dashed denote smooth transitions.  This system displays all three characteristic shapes; (b): By changing the value of $b$ so that the sheets do not come as close during adhesion, the tightly clamped region is seen to disappear; (c) and (d): For $p=3$ and $q=9$, further qualitative changes shrink the hysteresis region and cause the disappearance of characteristic shapes;  (e) and (f): In the case where $p=2$ and $q=4$, the hysteresis can disappear completely  even in the presence of a transition.}
\end{figure*}

Although our investigation to this point has considered particular values for the parameters $\{\beta, p,  q\}$,  qualitative changes in the hysteresis and transition behavior can be obtained  for different values of these parameters. Not only can the hysteresis region be made to shrink, but it can also disappear entirely.  In addition, while the transitions seen so far have been first-order, by tuning the model parameters this transition can be made to become second-order.

These different behaviors are illustrated in Fig.~\ref{energies}, where for convenience we have introduced the parameter $b=\beta J_p/J_q$.  As we saw above, there are three characteristic shapes for the sheets that we will denote as weakly bent (W), adhered/clamped (C), and arc-shaped (A).  The areas that exhibit each of these shapes are depicted in Fig.~\ref{energies}. In Fig.~\ref{energies}a and b, we display the total energy (bending plus attraction) for a system with $p=6$, $q=12$.  These values model the attraction between two thin non-polar filaments, and the characteristic shapes seen are similar to those in Fig.~\ref{hysteresis}.  If $b$ is increased,  representing an increase in the minimum adhesion distance between the fibers, the arc-shapes and associated phase transition vanish.  Similarly for the case where $p=3$ and $q=9$ \cite{oyhar}, increasing the value of $b$ causes the hysteresis region to shrink and the arc-shape vanishes (Figs.~\ref{energies}c and \ref{energies}d).  Furthermore, the hysteresis region can be made to disappear completely for  $p=2$ and $q=4$ (Fig.~\ref{energies}e and \ref{energies}f).

\begin{figure}[t]
\includegraphics[width=.5\textwidth]{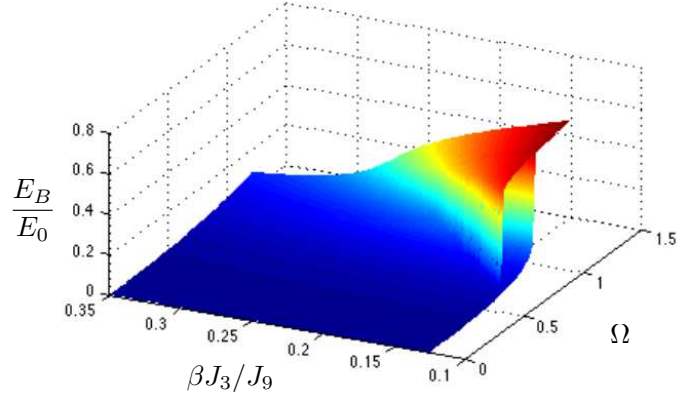}
\caption{(Color online) \label{cusp} Bending energy landscape for $p=3$, $q=9$ over a range of $\beta$ and $\Omega$.  Note that there is a distinct cusp in this parameter space, indicative of the ``catastrophic'' behavior that is associated with first order phase transitions. }
\end{figure}

The transformation of a first-order transition into a second-order transition indicates that there may be a cusp catastrophe in the parameter space we are exploring \cite{strogatz}.  For  the case $p=3$ and $q=9$, we plot on Fig.~\ref{cusp}  the bending energy landscape as both $\Omega$ and $b=\beta J_3/J_9$ are varied.  For small values of $\beta$, the first-order nature of the transition is apparent.  For a fixed value of $\Omega$, increasing $\beta$ decreases the area of hysteresis, until a critical value is reached where the hysteretic behavior vanishes completely.

\subsection{Non-identical sheets}

\begin{figure}[b]
\includegraphics[width=.5\textwidth]{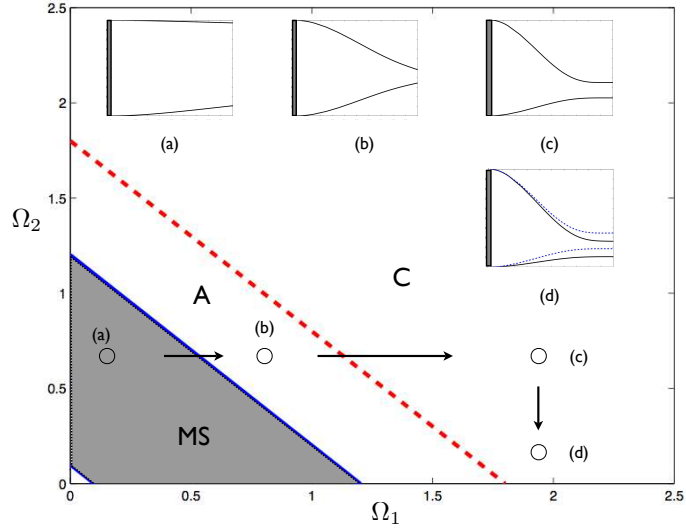}
\caption{(Color online) \label{nonsymmetric} Phase diagram of sheets with asymmetric bending parameter for $p=3$, $q=9$, and $\beta=0.15J_3/J_9$.  There is a region of multiple stability (MS) where arc-shapes and weak bending are both possible, arc-shapes only (A) and tightly clamped (C) sheets. (a): In this region,  the sheets are either bending weakly, or in the hysteretic case they exhibit arc-shapes;  (b): As $\Omega_1$ is increased, the sheets snap together into an arc-shape (a first-order transition);  (c) As $\Omega_1$ increases further, there is a smooth variation from arc-shapes into clamped ({\it i.e.} a second-order transition, dashed line); (d) If $\Omega_1$ remains fixed, and $\Omega_2$ is decreased, the bottom sheet will become relatively more rigid, producing a net shift in the equilibrium adhesion position (the dashed lines indicate the adhesion shape in Fig.~\ref{nonsymmetric}c).}
\end{figure}

We have considered so far the case where the two sheets are identical.  If we allow instead their bending rigidity   to be different, we now have two parameters,  $\Omega_i=(2WJ_pL^3\sigma^p)/(B_ih^p\epsilon)$ for $i=1,2$.  In this case,  the difference in rigidity causes a change in both the equilibrium point of adhesion and the critical values of $\Omega_i$.  The different shapes can be characterized by a phase diagram that maps the transition points for different values of the two bending parameters, as displayed on Fig.~\ref{nonsymmetric}.  Three distinct regions exist: Multistability of both arc-shapes and weak bending (MS), arc shapes exclusively (A), and adhered or clamped states (C).  As shown in Fig.~\ref{nonsymmetric}a, for small values of $\Omega_2$ and $\Omega_1$, weak attraction occurs, unless the hysteretic regime has been entered, in which case there will be adhesion (not shown).  As $\Omega_1$ is increased past the fixed value of $\Omega_2$, the sheets adhere in an asymmetric arc shape (Fig.~\ref{nonsymmetric}b).  This transition is first-order, and  the sheets snap together. As $\Omega_1$ crosses the second-order transition threshold (dashed line in Fig.~\ref{nonsymmetric}), the shapes smoothly evolve into a clamped phase (Fig.~\ref{nonsymmetric}c).  In this final clamped state, the  position of the adhered portion of the sheets depends on the relative values of the bending parameters (Fig.~\ref{nonsymmetric}d).

\subsection{Adhesion of three sheets}

We now consider the  adhesion transition for  an array of multiple sheets, and illustrate the complexity and richness of the system considered on a few examples. Assuming nearest-sheet interaction for simplicity, with identical potentials, we can easily extend the modeling approach offered above to the case of $N$ interacting sheets \footnote{Even for $p=1$ (long-ranged or Coulombic attraction) adding contributions from all neighbors does not qualitatively change our results. }. The equation of shape for the $i$th  sheet, $1\leq i \leq N$, is then given by
\begin{eqnarray}
y_i^{\prime\prime\prime\prime}+\Omega_i\sum_{j}\left[\frac{1}{(y_i-y_j)^p} -\frac{\beta}{(y_i-y_j)^{q}}\right]=0,
\end{eqnarray}
where the sum on $j$ runs over nearest neighbors, and $\Omega_i$, $p$, $q$, and $\beta$ are defined as in the $N=2$ case.  

\begin{figure}[t]
\includegraphics[width=.6\textwidth]{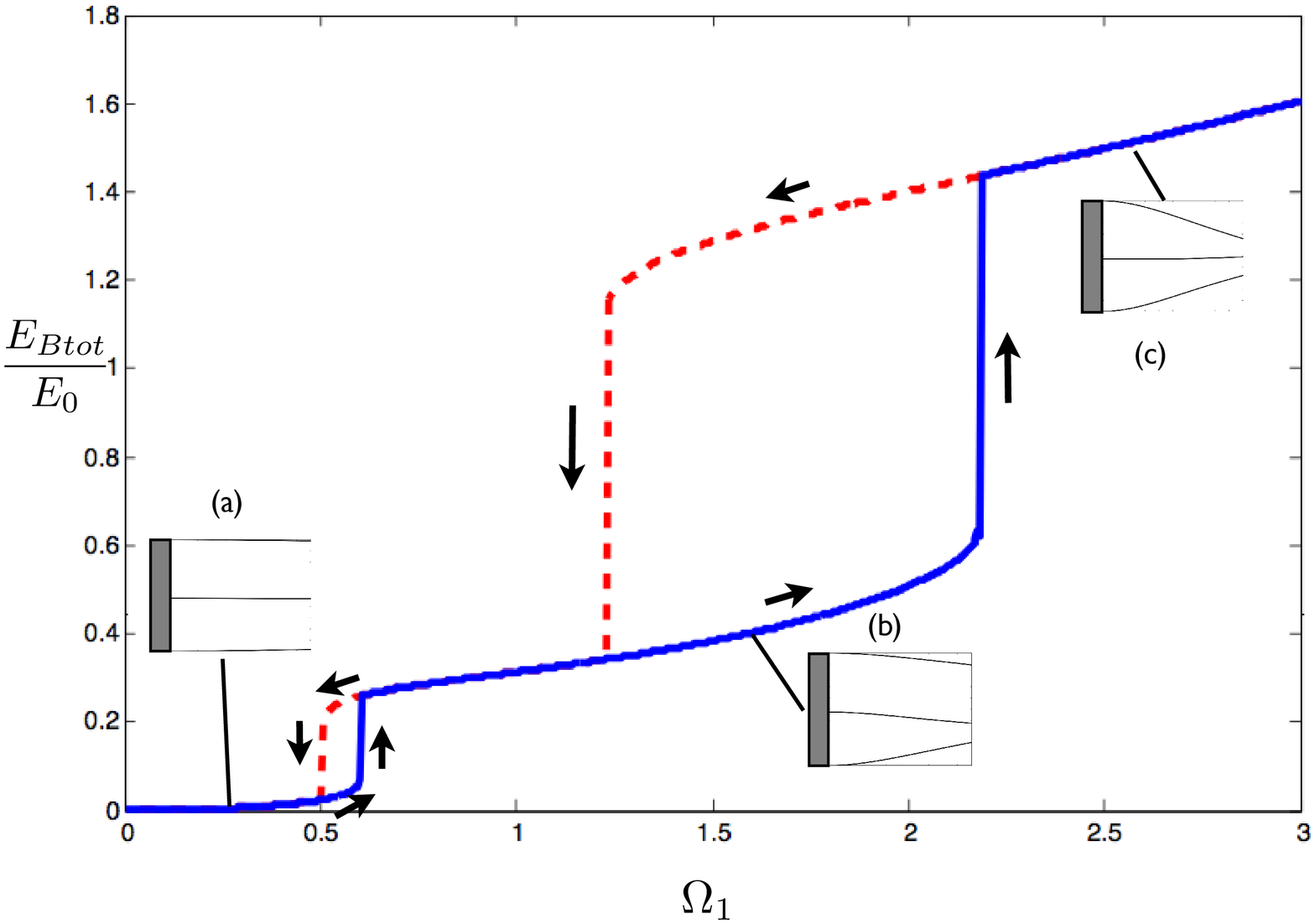}
\caption{(Color online) \label{threehyst} Sequential adhesion transition for three sheets: $(1,1,1)\to(1,2)\to(3)$. Two consecutive first-order transitions occurring for the total bending energy of the sheets as a function of the (identical) bending parameters  in the case  $p=3$,  $q=9$ and $\beta=0.35J_9/J_3$.  An inherent asymmetry has been introduced in that the middle sheet is pinned slightly closer to the bottom sheet than to the top one (five percent difference in height).  (a):  Weak attraction; (b) 
As the bending parameters are increased,  a first-order adhesion transition takes place where two of the three sheets adhere; (c): A second first-order transition occurs when the three sheets adhere. Both transitions display hysteresis.}
\end{figure}

\begin{figure}[b]
\includegraphics[width=.6\textwidth]{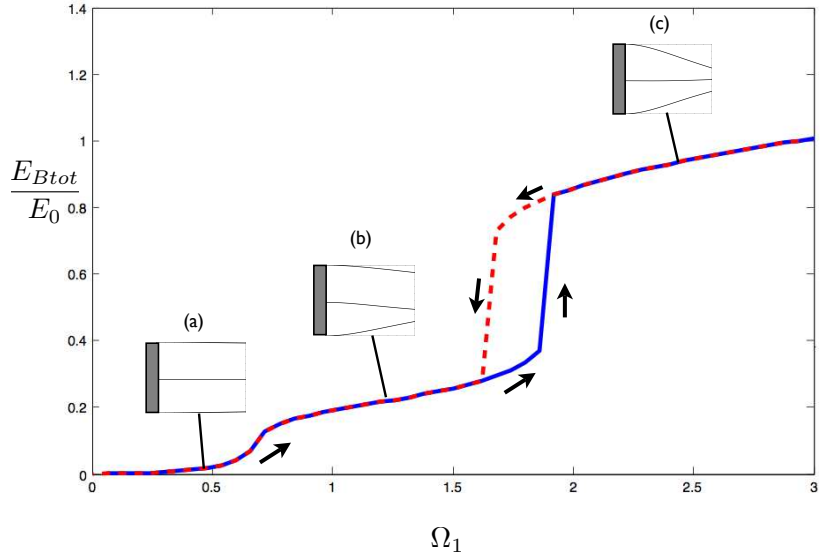}
\caption{(Color online) \label{threehyst2} Same as Fig.~\ref{threehyst}, but for $\beta=0.45J_9/J_3$. In contrast with the case depicted in Fig.~\ref{threehyst}, the first adhesion transition is now of second order.}
\end{figure}

For multiple sheets, any asymmetry in the system now plays a role in determining the order in which sheets adhere to one another. For three sheets, unless there is perfect symmetry between the top and bottom sheet, the adhesion events always occur in a sequential fashion (see Fig.~\ref{threehyst} and \ref{threehyst2}). Specifically, two sheets first come together, and then adhere to the third sheet for a further increase in the relevant bending parameter. By locating the middle beam slightly closer to one of its neighbors, this sequential can be made  to occur preferentially between two previously-chosen sheets (by changing the clamping distance, the competition between bending and interaction energy changes, and in effect one of the bending parameters gets a boost from the geometric asymmetry).   For three identical sheets, as the bending parameters are increased, there is a transition from the weakly attracted phase (Fig.~\ref{threehyst}a and \ref{threehyst2}a) to a regime where two  sheets adhere to each other (Fig.~\ref{threehyst}b and \ref{threehyst2}b).  As the bending parameters are further increased,  a subsequent transition occurs where all three sheets come together (Fig.~\ref{threehyst}c and \ref{threehyst2}c).  As for the $N=2$ case, tuning the value of $\beta$ can change the nature of the first adhesion transition (a$\to$b), from  first-order (Fig.~\ref{threehyst}) to  second-order (Fig.~\ref{threehyst2}).  The second adhesion transition, however, remains first-order.

\subsection{Adhesion of four sheets}

\begin{figure}[t]
\includegraphics[width=.6\textwidth]{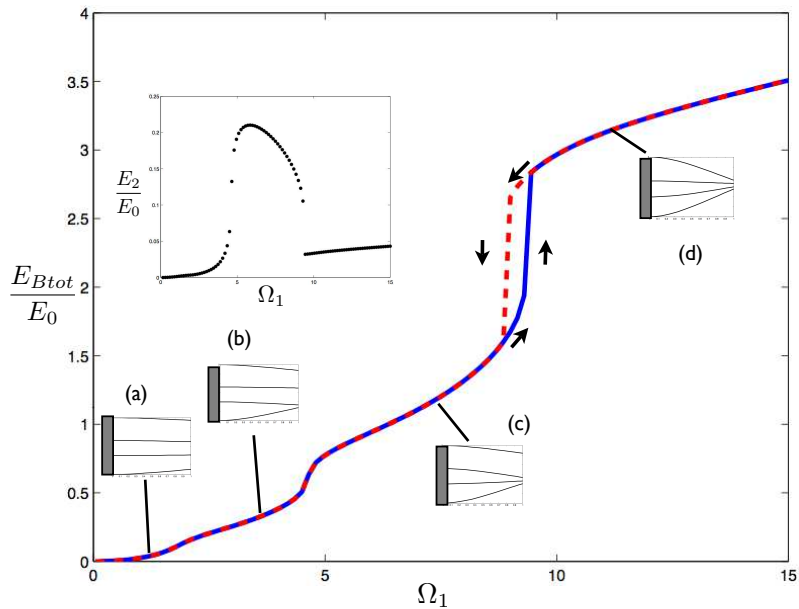}
\caption{(Color online) \label{1_3energy} Sequential adhesion transition for four  sheets: $(1,1,1,1)\to(1,1,2)\to(1,3)\to(4)$. Total bending energy as a function of the (identical) bending parameters for $p=1$, $q=2$, and $\beta=0.1J_2/J_1$. The four sheets form first an adhered pair, then a triplet, then all four clamp together.  Adhesion in this manner is highly dependent on the asymmetry of the array of multiple structures.  In this case the top sheet is placed at $y=3h$, the second highest at $y=1.8h$, the third at $y=0.7h$ and the lowest beam placed at $y=0$.  (a): Weakly bent state; (b): Adhesion between the lower two sheets; (c): Adhesion between the lower three sheets; (d): Adhesion of all four sheets. Inset: Non-monotonic variation of the bending energy in the second-highest sheet.}
\end{figure}

\begin{figure}[t]
\includegraphics[width=.6\textwidth]{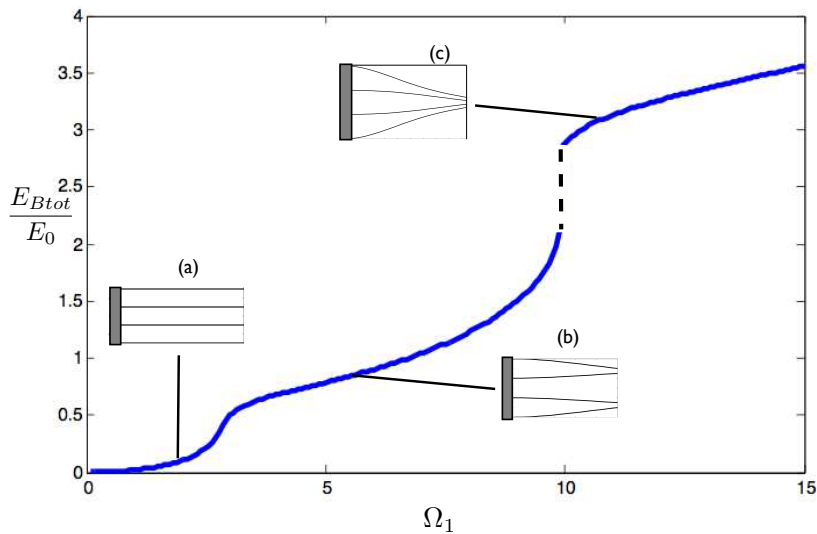}
\caption{(Color online) \label{2_2beam} Same as Fig.~\ref{1_3energy} except that the original distances between each sheet is now identical. As a result, the adhesion transitions  occur in a hierarchical fashion: $(1,1,1,1)\to(2,2)\to4$.}
\end{figure}

When a fourth sheet is added, the adhesion transitions can be made to occur in either a hierarchical  or a  sequential fashion. This is illustrated in Figs.~\ref{1_3energy}  and \ref{2_2beam}. 

We show in Fig.~\ref{1_3energy}  an example of sequential adhesion for four identical sheets, similar to the one discussed in the three-sheet case.  The values of $p$, $q$ and $\beta$, as well as the geometric asymmetry, have been chosen so that there is a mix of first- and second-order transitions (see figure captions).  The relative distances between the four sheets are as follows: The top sheet is pinned at $y=3h$, the second-highest at $y=1.8h$, the third-highest at $y=0.7h$ and the final sheet at $y=0$.  As all four bending parameters are increased at the same rate, the sheets start by  a state of weak attraction, with more bending exhibited by the sheets that are closer to one another.  Past a critical value of the bending parameters, there is a second-order transition for these values of the model parameters, and the two bottom sheets adhere 
(Fig.~\ref{1_3energy}b).  As the bending parameters are increased further, another second-order transition takes place and the lower three sheets adhere (Fig.~\ref{1_3energy}c).  Finally, a final first-order transition occurs when the four sheets adhere (Fig.~\ref{1_3energy}d). Note the non-monotonic variation of the bending energy in the second-highest sheet (inset of Fig.~\ref{1_3energy}).

An example of hierarchical adhesion transition is displayed in Fig.~\ref{2_2beam}, where we plot the total bending energy profile for a symmetric four-sheet system ({\it i.e.} there is no asymmetry in the relative distances between the sheets).  As the bending parameters are increased, a second-order phase transition leads to adhesion between two pairs of sheets (Fig.~\ref{2_2beam}a).  First-order transitions are also possible for other values of the model parameters (not shown here).  As the bending parameters are increased further,  a first-order transition occurs and the four sheets all adhere to one another (Fig.~\ref{2_2beam}b).   Remarkably, in this case, the first-order transition does not display any hysteresis.

\section{Conclusion}

In this paper we have studied the prototypical  dry adhesion problem between flexible sheets or filaments, and focused on their morphological transitions. Motivated by a simple macro-scale experiment showing hysteretic adhesion, we have introduced a model of dry adhesion between two elastic, slender sheets interacting via a power-law potential, and studied numerically the transitions in their  conformations.  Given a particular form of interaction potential,  the system is  completely described by a single dimensionless parameter quantifying the relative effect of long-range attraction and bending rigidity, and governing the nature of the adhesion transitions (first or second-order). We have also generalized the model to multiple sheets, showing in particular that additional geometric considerations dictate the order in which structure adhere to each other. The physical systems modeled here include the interactions between charged sheets, or between nonpolar filaments. 
Future work will focus on the presence of thermal fluctuations allowing the adhered states to ``jump'' from one state to another. We will also consider the case where the filaments are actuated, and will include the effect of hydrodynamic interactions.  Finally, using an approach similar to ours, the adhesion of three-dimensional structures such as coiled filaments or planar arrays could be investigated.

\section*{Acknowledgments}
We would like to thank Saverio Spagnolie for many useful conversations and help with some numerical aspects of this work.  This work was funded in part by the National Science Foundation (grants CTS-0624830 and CBET-0746285 to Eric Lauga).

\bibliographystyle{unsrt}
\bibliography{adhesion_paper.bib}

\end{document}